\documentclass[twocolumn,showpacs,showkeys,amsmath,amssymb]{revtex4}

\usepackage{graphicx}% Include figure files
\usepackage{dcolumn}% Align table columns on decimal point
\usepackage{bm}% bold math

\begin{document}

\title{Proton vs. neutron halo breakup}% Force line breaks with \\

\author{Angela Bonaccorso}
\email{bonaccorso@pi.infn.it}
\affiliation {Istituto Nazionale di Fisica Nucleare, Sezione di Pisa, 56127
  Pisa, Italy}
\author{  David M. Brink}%
\email{thph0032@herald.ox.ac.uk}
\affiliation{Department of Theoretical Physics, 1 Keble Road, Oxford OX1 3NP, U.K.}
\author{Carlos A. Bertulani}
\email{bertulani@nscl.msu.edu}
\affiliation{NSCL, Michigan State University, East Lansing, MI 48824, USA.}%

\date{\today}% It is always \today, today,

\begin{abstract}
In this paper we show how effective parameters such as effective binding energies can be defined for
a proton in the combined nuclear-Coulomb potential, including also the target potential, in
the case in which the proton is bound in a nucleus which is partner of a nuclear reaction. Using such
effective parameters the proton behaves similarly to a neutron. In this way   some unexpected
results obtained from dynamical calculations for  reactions initiated by very
weakly bound proton halo nuclei can be interpreted. Namely the fact that stripping dominates the nuclear
breakup cross section which in turn dominates over  the Coulomb breakup even when the target is
heavy at medium to high incident energies. Our interpretation helps also clarifying why the existence 
and characteristics of a proton halo extracted from different types of  data have sometimes appeared
contradictory.
\end{abstract}


\maketitle

\section{\bf I. INTRODUCTION.}

This paper is concerned with the differences which might arise in
reactions initiated by a neutron halo nucleus like $^{11}$Be and a
proton halo nucleus like $^{17}$F or $^{8}$B. Halo nuclei are a
special case of radioactive beams for which the last nucleon is
very weakly bound, with separation energies of the order of 0.5
MeV or less, and in a state of low angular momentum (l=0,1). They
exhibit extreme properties like very large total and breakup cross
sections. Nuclear and Coulomb breakup of neutron halo nuclei have
been studied in great detail both experimentally as well as
theoretically and are now quite well understood processes
\cite{hs}. On the other hand proton halo nuclei such as $^{8}$B
and $^{17}$F are still under investigation.  Their behavior as
projectiles of nuclear reactions needs to be understood better in
particular as $^{8}$B is  partner in (p,$\gamma$) radiative
capture reactions of great astrophysical interest for the
understanding of the neutrino flux from the sun (see for example
the discussion and references of \cite{dave}). Also the existence of
 a proton halo has sometimes been questioned \cite{kelly} and
results from different experiments might seem to be contradictory \cite{gai}.
 For those nuclei Coulomb breakup reactions in the
laboratory have been used to get indirect information on the radiative capture, since it has been
shown that the Coulomb breakup cross section is proportional to
the radiative capture cross section \cite{ber}.

 In the case of neutrons the Coulomb breakup cross section is largest for heavy targets and the
interplay with nuclear  breakup is well understood both experimentally as well as
theoretically, in particular thanks to the measurements of angular distributions for both
processes \cite{anne,nak94, bon98a,mar02}. Then
$^{208}$Pb and
$^{58}$Ni have  been used as targets with beams of $^{8}$B or of $^{17}$F at various energies
\cite{dave}\cite{mot}-\cite{blan}.
Data on lighter targets such as $^{9}$Be and $^{28}$Si \cite{ kelly, borcea,neg}
also exist. At the same time a  number of theoretical papers have appeared dealing with the problem of the
accuracy necessary to interpret the data \cite{fil}-\cite{tt}. In particular the problems of
higher order effects in Coulomb breakup, of the inclusion of E0, E1, and E2 multipolarities in the Coulomb field
and of the relative magnitude of nuclear and Coulomb contributions and of their interference have been discussed
at length.

A number of experimental papers \cite{neg,blan} have
shown that for a $^{8}$B projectile it is the nuclear breakup and in particular the stripping (or absorption)
\cite{borcea,neg} component that dominates the experimental cross section. In \cite{borcea,neg} a $^{28}$Si target was 
used and different beam energies around 40A.MeV were explored. The data of Table I of \cite{neg} show
that stripping (110$\pm$ 9 mb) is very close to the total diffraction (112$\pm$ 12mb) which contains both nuclear and
Coulomb components. On the other hand at the same beam energy and on the same target the one neutron breakup of
$^{11}$Be, measured by \cite{neg2} and calculated by \cite{abfc} gave a stripping cross section of 220 mb and a total
diffraction of 300 mb of which 120 mb from Coulomb breakup. These results could be considered
rather astonishing in view of the fact that the proton in $^{8}$B has a separation energy of
0.14 MeV while the neutron separation energy in 
$^{11}$Be is larger and equal to 0.5 MeV. On the other hand the data of \cite{blan} for the breakup of $^{8}$B on
$^{208}$Pb at 142A.MeV provided a  one-proton removal cross section of 744$\pm$9 mb of which about 300-450 mb were
estimated to be due to nuclear breakup and 311 mb to Coulomb breakup.  This is again a surprising result because for
the system
$^{11}$Be+$^{208}$Pb at 120A.MeV it was calculated in \cite{mar02}  that the cross sections would be 321 mb for nuclear
breakup and 1050 mb for Coulomb breakup, the model of \cite{mar02} being very reliable as it agrees with exclusive
data \cite{anne}. Similarly, the recent data from GSI \cite{lola} at the relativistic beam  energy of 936 A.MeV give for
the one proton removal cross section of $^{8}$B on $^{208}$Pb and $^{12}$C, 662$\pm$60 mb and 94$\pm$8mb respectively
while at a similar energy (790A.MeV) the one neutron removal from $^{11}$Be on the same targets was 960$\pm$60mb and 
169$\pm$4 respectively \cite{kob}.

On the other hand, very recently a new theoretical  work has appeared where the authors treat the nuclear breakup of
$^{17}$F to first order
\cite{berpaw}, contrary to what it has been established in the literature, namely that halo breakup should
be treated to all orders in the neutron-target interaction. An a posteriori justification of the approach
of Bertulani and Danielewicz
\cite{berpaw} is that the calculated nuclear breakup is larger by several orders of magnitude than the
Coulomb breakup.  In fact the approach
of
\cite{berpaw} can be justified with the results of another theoretical work by Esbensen and Bertsch
\cite{esbber1} on the proton halo nucleus $^{8}$B, where it was shown that starting from
about 40 A.MeV in the reaction $^{8}{\rm B}+^{208}{\rm Pb}$, dynamical calculations and
first order perturbation theory with or without far field approximation, yields nearly
    the same  Coulomb breakup cross sections for  distances of closest approach  for the core target trajectory of
20 fm or larger. Also in \cite{esbber} the same authors found that for $^{17}$F nuclear
diffraction and Coulomb breakup have very similar probabilities to occur and the values are
also close to those for  nuclear stripping. An earlier calculation by Esbensen and Hencken \cite{esbhenk} showed that
nuclear one proton removal   cross sections for a $^{8}$B projectile would be larger than Coulomb cross sections up
to target mass A$_T$=100.  Similar conclusions were reached by Dasso, Lenzi and Vitturi
\cite{DLV}. 

In order to get some insight into the peculiarities of the proton halo  reactions, in
particular in comparison to neutron halos, we introduce here an effective treatment of proton
single particle states which simplifies their treatment and makes them behaving as neutrons.
Related approaches have recently been introduced by other authors \cite{carmen}. We do not
propose our method  as opposed to dynamical
calculations, but we are simply concerned with the understanding of the underlying  physics and
the interpretation of  numerical results from more sophisticated methods such as direct solutions of the Schr\"odinger
equation or coupled channels. 

We show in the following how to treat proton transfer and breakup in a way that is similar to
neutron transfer and breakup by using an effective potential in which the weakly bound
protons behave as ``normally" bound neutrons and then we come to some simple  conclusions.
The basic idea is that breakup is a kind of ``transfer to the continuum" and as such its
main features come from matching conditions and Q-value effects \cite{bon99}.

\section {Proton vs. neutron: effective potential}
\begin{figure}[pht]
\center
\includegraphics[scale=0.4, angle=90]{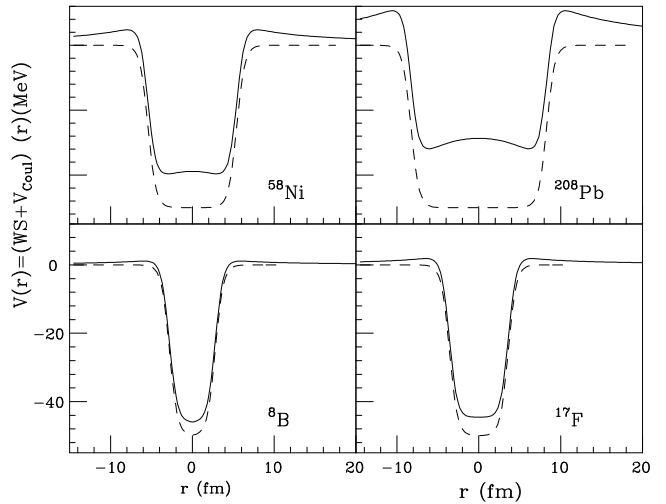}
\caption{Nuclear (dashed) and nuclear-Coulomb (solid) potentials for $^{8}B$, $^{17}F$, $^{58}Ni$ and
$^{208}Pb$. }
\label{fig1}
\end{figure}

We begin this section by noticing the differences in the treatment of a neutron halo breakup and a
 proton halo breakup. In ref.\cite{mar02} the neutron breakup was studied to all orders in the nuclear and Coulomb
fields. The nuclear potential responsible for the neutron transition to the continuum was taken to be the
neutron-target optical potential. On the other hand it was shown that Coulomb breakup  originates from an effective
repulsive force acting on the neutron and due to the core-target Coulomb potential. If we were to extend the same
model to proton breakup we should add the two Coulomb interactions of the proton itself with its core in the initial
state and with the target in the final state. Now, because of the slow variation of the Coulomb field,  we can use the
 adiabatic approximation or frozen halo  \cite{rj}, for these two Coulomb interactions which make the proton breakup
different from the neutron breakup. The detailed derivation of the formalism is given in the Appendix.

 The effect of the proton-core and proton-target Coulomb potentials can be understood qualitatively by discussing
Fig. 1 which shows the potentials felt by a neutron (dashed line) and a proton (full line) in
$^{8}$B,
$^{17}$F,
$^{58}$Ni and
$^{208}$Pb.
Supposing the two particles have the same binding energy $\varepsilon_i<0$
the proton wave function inside the potential of Fig. 1 is  like
a neutron wave function with binding energy $\varepsilon_i-Z_Pe^2/R_i$
up to the radius $R_i$. The proton potential is  like the neutron
potential pushed up by $Z_Pe^2/R_i$ where $R_i$ is the barrier
 radius. For any given nucleus this radius is
rather larger than the nuclear or Coulomb radius  values usually quoted in the literature.
But from  Fig. 1 one can see that it is the value corresponding to the
barrier peak. We give these values in Table I, together with the experimental binding energies of the
halo state in $^{8}$B  and of two states in $^{17}$F .

\begin{table}[h]
\caption{Barrier radii from Fig. 1 and initial binding energies.}
\vskip.3in \begin{center}
\begin{tabular}{lccccccc}
\hline\
                     & $^{8}{\rm B}$&J$^\pi$& $^{17}{\rm F}$&J$^\pi$&$^{58}{\rm Ni}$&$^{208}{\rm Pb}$\\ \hline
$R_{i,f}(fm)$ & 6.0 && 6.5&&8.0 &10.5\\
$\varepsilon_i(MeV)$ & -0.14 &$1p_{3/2}$    & -0.6 &$1d_{5/2}$&--&-- \\
$\varepsilon_i^*(MeV)$ &-- & &-0.1 & $2s_{1/2}$&--&-- \\
\hline
\end{tabular}\end{center}
\end{table}

\begin{table}[h]
\caption{Effective parameters.}
\vskip.3in \begin{center}
\begin{tabular}{lccccc}
\hline\
                    & & $^{8}{\rm B}+^{58}{\rm Ni}$ &$^{8}{\rm
B}+^{208}{\rm Pb}$&$^{17}{\rm F}+^{58}{\rm Ni}$&$^{17}{\rm F}+^{208}{\rm Pb}$\\
\hline
$\Delta_{i}(MeV)$ &&  -1.85   & -2.29 & -2.7& -3.2 \\
${\tilde\varepsilon}_{i}(MeV)$ && -1.99   & -2.43 &-3.3&-3.8 \\
${\tilde\gamma_i}(fm^{-1})$ && 0.29  & 0.34 &0.39&0.42 \\
${\tilde C_i}(fm^{-1/2})$ && 0.69  & 0.79 &0.75&0.89 \\
${\tilde\varepsilon}_{i}^*(MeV)$ && --    & --&-2.8&-3.3 \\
${\tilde\gamma_i}^*(fm^{-1})$ && -- & -- &0.36&0.39 \\
${\tilde C_i}^*(fm^{-1/2})$ &&--  & -- &3.06&3.5\\
\hline
\end{tabular}\end{center}
\end{table}

\begin{figure}[pht]
\center
\includegraphics[scale=0.4, angle=90]{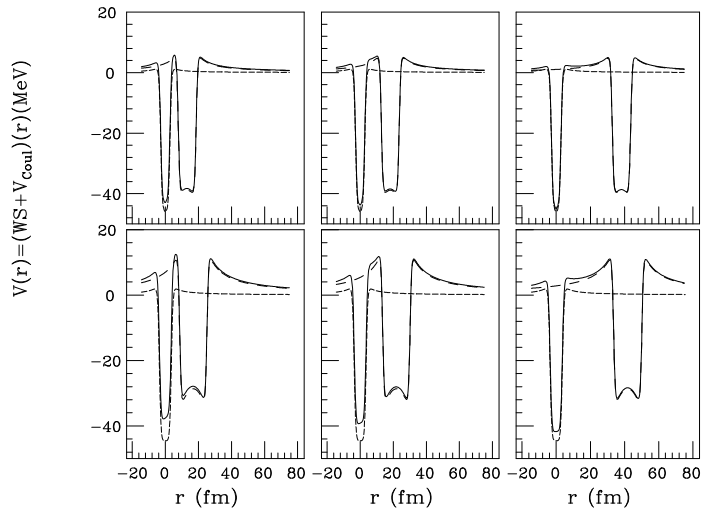}
\caption{Nuclear-Coulomb  potentials for $^{8}{\rm B}+^{58}{\rm Ni}$ (top) and  $^{17}{\rm F}+^{208}{\rm Pb}$
(bottom) at
 distances between the centers equal to $d=1.4(A_p^{1/3}+A_T^{1/3})\ {\rm fm}+s$, with $s=5$, 15 and 30 fm. Short
and long dashed lines are the  projectile and target potentials respectively. Full line is the projectile-target
combined  potential. }
\label{fig2}
\end{figure}

But as it is shown in the Appendix, in a scattering process there is also an effect due to the Coulomb
potential of the projectile.
It can be understood by looking at Fig. 2 which shows the nuclear-Coulomb  potentials for
$^{8}{\rm B}+^{58}{\rm Ni}$ (top) and  $^{17}{\rm F}+^{208}{\rm Pb}$ (bottom) at several distances. Short and long
dashed lines are the separate projectile and target potentials respectively. Full line is the
projectile-target combined  potential.
The  effect of the target potential on the projectile potential is actually twofold:
\begin{itemize}
\item {} the center of the projectile potential shifts up by an amount $Z_Te^2/d$
where $d$ is the distance of closest approach between the two nuclei.

\item {} the height of the barrier on the side near the target goes up
by an amount $Z_Te^2/|d-R_i|$ relative to the center. While on the other side it goes up by
$Z_Te^2/|d+R_i|$. 
\end{itemize}
 This suggests
that the true binding energy $\varepsilon_i$ could be
 replaced by 

\begin{equation} \varepsilon_i\to {\tilde\varepsilon}_i= \varepsilon_i-\Delta_i.\label {1a}\end{equation}

Where \begin{equation}\Delta_i={Z_Pe^2\over R_i}
+Z_Te^2 \left({1\over
2}\left({1\over{|d+R_i|}}+{1\over{|d-R_i|}}\right)-{1\over d} \right).
\label{4a}\end{equation}

${Z_Te^2\over
2}\left({1\over{|d+R_i|}}+{1\over{|d-R_i|}}\right)$ is the average effect of the target Coulomb potential 
at the points
$r=\pm R_i$ on the left and right sides of the projectile. 

In the reactions we are discussing the initial states
are always  bound. According to Eq.(\ref{1a}) they will be shifted down by a $\Delta_i$.  
Therefore the phase space for breakup states will be reduced 
and thus breakup probabilities for protons will be smaller than for neutrons having the same binding energy. 
Furthermore there will be an important target dependence.

Then we conclude that some features of  proton  breakup could be understood by analogy
with   neutron breakup by using effective parameters in the following way:
\begin {itemize}

\item {} use effective $\tilde \gamma_i $ 
calculated from
\begin{equation}{\hbar^2\tilde \gamma_{i}^2\over 2m}=|{\tilde\varepsilon}_{i}|\end{equation}

\item {} calculate the normalization constants $\tilde C_{i}$ 
 of asymptotic wave functions \cite{picc} as for neutron wave functions with binding energies
${\tilde\varepsilon}_{i}$.
\end {itemize}
The approach here corresponds to an adiabatic approximation for the effect
of the Coulomb force of one nucleus on the other   and it
was introduced for the first time in Ref.\cite{9} where it was also shown that it is
equivalent to the sudden approximation which was instead discussed in Ref.\cite{10}.
We used it already in \cite{2} to discuss the proton transfer to the continuum reaction
$^{197}{\rm Au}(^{20}{\rm Ne},^{19}{\rm Fl})^{198}{\rm Hg}$ \cite{wald}. The approximation of using effective
parameters for protons so that they could be treated similarly to neutrons has a long story in direct
reaction theories, see for example
\cite{bw}. However it is worth noticing that the definitions Eqs.(\ref{1a})  used here
represent a generalization and improvement with respect to those used in \cite{9,10,bw}. This
is because we not only take into account the effect of the Coulomb barrier in the projectile
and target potentials, but also consider the "polarization" effect that the target Coulomb
potential has on the projectile and  vice versa.
 As Fig. 2 (bottom
part) shows, in the case of a light projectile and a heavy target, the long range effect of the Coulomb
potential gives a considerable shift upwards of the projectile potential. The importance of such effects and the
meaning of the approach discussed here will be clearer later on when we will discuss Fig. 3 for the wave
functions.

 From Fig. 2 one sees clearly that the effect of the barrier is very
important even at distances as large as 30 fm. In order to
quantize the effects discussed above we give in Table I the
barrier radii, called $R_i$ and $R_f$, for two nuclei usually used
as projectiles $^{8}{\rm B}$, $^{17}{\rm F}$, and for $^{58}{\rm
Ni}$, $^{208}{\rm Pb}$, which have been used as targets. In this case $R_f$ has been taken as the barrier radius which
can be deduced from Fig. 1. In Table II we give  the effective energy  shift $\Delta_i$, the
effective binding  energies  for the possible projectile-target combinations discussed in this
paper. For completeness we add the effective length parameters $\tilde \gamma_i$ and asymptotic
normalization constants $\tilde C_i$ of the initial
 asymptotic wave functions. It is indeed the tail of the wave function which determines the main
characteristics of the breakup mechanism (\cite {bbst} and references therein).
\begin{figure}[pht]
\center
\includegraphics[scale=0.4, angle=90]{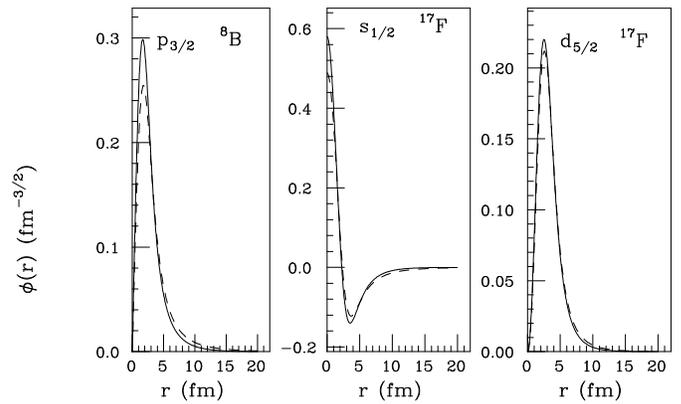}
\caption{Proton (dashed) and neutron (solid) wave functions for $^{8}{\rm B}$, $^{17}{\rm F}$ as indicated. Neutron
wave functions obtained for effective energies as in Table II, in the case of the $^{58}{\rm Ni}$ target.}
\label{fig3}
\end{figure}

We illustrate the last point by Fig. 3 where  the dashed lines   represent the
proton single particle wave functions corresponding to the three   initial states of
Table I, calculated in a Woods-Saxon plus spin-orbit \cite{bbst} plus Coulomb potential
with parameters:
$r_0=1.27fm~~~a=0.65fm~~~V_{so}=7MeV~~~r_c=1.3fm$. The Woods-Saxon depth is fitted to give the
correct  binding energy. The solid lines represent the neutron wave functions calculated
with the effective binding energies in the case of the $^{58}{\rm Ni}$ target. One sees clearly that in each case the
true proton wave function is very close  to the ``effective energy" neutron wave function.  We remind the reader that
at small distances the breakup is strongly reduced due to the core-target absorption into more
complicated reaction channels.

From the values  shown one clearly sees that the proton
halo behaves in a breakup reaction with a heavy target as a neutron state bound with a
``normal" energy of several MeV, for which it is very well known that the nuclear breakup
is comparable  to the Coulomb breakup and on the other hand that the stripping is
  dominant on  diffraction \cite{end,in,hl}. 

To give an idea of
the orders of magnitude involved, we have calculated total breakup cross sections for two
reactions:
$^{11}{ \rm Be}\to ^{10}{\rm Be}+n$ and $^{17}{ \rm F}({1/2}^+)\to ^{16} {\rm O}+p$, both at 40 A.MeV on a
$^{208}{\rm Pb}$ target.
Nuclear and Coulomb breakup of $^{11}{ \rm Be}$ have been studied in many experiments on heavy targets  and
absolute breakup cross sections are very well known
\cite{anne,nak94}. For Coulomb breakup we used first order perturbation theory and for nuclear breakup we used
the transfer to the continuum model \cite{bon99,2}. Our aim here is only to give some order of magnitude
estimates. For the breakup of
$^{17}{\rm F}$  we used a neutron wave function and the ``effective parameters" of Table II. The values obtained are
given in Table III. In the case of $^{11}{ \rm Be}$ we have used a spectroscopic factor ${\rm C^2S}=0.77$ for the
initial state, while for $^{17}{ \rm F}$ we have used a unit spectroscopic factor. One sees that for the proton
``halo" state in 
$^{17}{
\rm F}$ there is a  strong reduction
 of  about a factor seven 
for Coulomb breakup and of a factor four for diffraction because both require the neutron to be in a final free
particle state, which is obviously less probable the stronger the ``effective binding" of the nucleon in the initial
state. For stripping instead the reduction is just a factor three. It is interesting to note that the reduction in the
proton removal cross sections from
$^{17}{\rm F}$ as compared to
$^{11}{\rm Be}$ calculated here and in \cite{berpaw} would be stronger than the reduction already seen in the 
data for $^{8}{\rm B}$ discussed in the Introduction. This is because $^{17}{\rm F}$ has a larger $Z_P$ than
 $^{8}{\rm B}$ and therefore as shown in Fig. 2 and Table II its effective separation energies are larger.

\begin{table}[h]
\caption{Cross sections in mb.}
\vskip.3in \begin{center}
\begin{tabular}{lcccc}
\hline\
                     && $\sigma_C$ &$\sigma_S$&$\sigma_D$\\ \hline
$^{11}{\rm Be}+^{208}{\rm Pb}$&&2724& 312  & 240 \\
$^{17}{\rm F~ }+^{208}{\rm Pb}$ &&382 &  131   & 53 \\
\hline
\end{tabular}\end{center}
\end{table}

 It appears
also clear that under such conditions Coulomb and nuclear breakup could not  need to be
calculated to all orders.  Also the effect of the ``effective parameters"  introduced here has
to be studied in more detail and results should be compared to full dynamical calculations.

\section{Conclusions}
In this paper we have tried to draw the attention 
to the physical origins  of the differences in the behavior in a reaction
of a proton halo nucleus as compared to a neutron halo. We have shown that
if the target is heavy, but also if the projectile is heavier,  as in the
case of
$^{17}{\rm F}$ vs.
$^{8}{\rm B}$ there is an effective barrier which makes the proton ``effectively" bound by several MeV, so that some typical halo
features might change in breakup reactions.
 In particular nuclear breakup and its stripping component could be of comparable magnitude as Coulomb breakup. 
This could explain the apparent discrepancy in the interpretation in terms of halo structure between   data from
different types of experiments. Also first order calculations are not completely unjustified. Therefore approaches of
the type used in
\cite{berpaw} although not very accurate would give reasonable order of magnitude predictions for weakly
bound protons interacting with a heavy target but not for interactions with light targets or in the case of neutron
breakup.

It is known that Coulomb breakup on a heavy target can be useful to simulate
the (p,$\gamma$) reactions of astrophysical interest. However, exclusive measurements need to be done to separate Coulomb from nuclear breakup. 
 Measuring
proton angular distributions as done in
\cite{anne} for neutron would help disentangling the dominant reaction mechanism, but also separating the large
core-target impact parameter contributions as done in
\cite{dave,nak94} is very useful.

 {\bf Acknowledgments}

This work was initiated while one of us (A.B) was visiting the Theory Group of the National Superconducting
Cyclotron Laboratory at Michigan State University. She wishes to thank  in
particular P. Gregers Hansen,  Pawel Danielewicz, Betty Tsang and Bill Lynch   for the very warm and stimulating
atmosphere and the hospitality extended to her.

 \appendix \section{Coulomb potentials}

We consider the breakup of a proton halo nucleus like $^{17}$F consisting of a
proton bound to a $^{16}$O core in a collision with a target nucleus.  The system of the halo nucleus and the target is
described by Jacobi coordinates $\left(  \mathbf{R,r}\right)  $ where $\mathbf{R}$ is
the position of the center of mass of the halo nucleus relative to the target
nucleus and $\mathbf{r}$ is the position of the neutron relative to the halo
core, and the coordinate $\mathbf{R}$ is assumed to move on a classical path.
This allows target recoil to be included in a consistent way. The Hamiltonian of the system is%

\begin{equation}
H=T_{R}+T_{r}+V_{pc}\left(  \mathbf{r}\right)  +V_{2}\left(  \mathbf{R,r}%
\right)  \label{h1}%
\end{equation}
where $T_{R}$ and $T_{r}$ are the kinetic energy operators associated with the
coordinates $\mathbf{R}$ and $\mathbf{r}$ and $V_{pc}$ is the potential
describing the interaction of the proton with the core, and it contains nuclear and Coulomb parts. The potential $V_{2}$
describes the interaction between the projectile and the target. It is a sum
of two parts depending on the relative coordinates of the proton and the
target and of the core and the target%
\begin{equation}
V_{2}\left(  \mathbf{R,r}\right)  =V_{pt}\left(  \mathbf{\beta}_{2}%
\mathbf{r+R}\right)  +V_{ct}\left(  \mathbf{R-\beta}_{1}\mathbf{r}\right)
\label{pot1}%
\end{equation}
Here $\beta_{1}=m_{p}/m_{P}$, $\beta_{2}=m_{c}/m_{P}=1-\beta_{1}$, where
$m_{p}$ is the proton mass, $m_{c}$ is the mass of the projectile core and
$m_{P}=m_{p}+m_{c}$ is the projectile mass. Both $V_{pt}$ and $V_{ct}$ are
represented by complex optical potentials. The imaginary part of $V_{pt}$
describes absorption of the proton by the target to form a compound nucleus.
It gives rise to the stripping part of the halo breakup. The imaginary part of
$V_{ct}$ describes reactions of the halo core with the target. The potentials $V_{pt}$ and
$V_{ct}$ also includes the Coulomb interaction between the proton and the target and the halo core and the
target. This part of $V_{ct}$ is responsible for Coulomb breakup.

The mass ratio $\beta_{1}$ is small for a halo nucleus with a heavy core. For
example $\beta_{1}\approx0.06$ and $\beta_{2}\approx0.94$ in the case of $^{17}%
$F. This property is used here to approximate the proton-target and
core-target potentials by

\begin{eqnarray}
V_{pt}\left(  \mathbf{\beta}_{2}\mathbf{r+R}\right)&  \approx&
V_{pt}(\mathbf{r+R)}\label{vnt}\\
V_{ct}\left(  \mathbf{R-\beta}_{1}\mathbf{r}\right)& \approx&
V_{ct}(\mathbf{R})+\mathbf{V}_{eff}\left(  \mathbf{r,R}\right)  \label{vnc}%
\end{eqnarray}

 The halo breakup is caused
by the direct proton target interaction $V_{pt}$ or by a recoil effect due to
 the core-target interaction. Coulomb breakup of a one-proton halo nucleus
is mainly a recoil effect due the Coulomb component $V_{ct}$ of the core-target
interaction and is contained in $\mathbf{V}_{eff}\left(  \mathbf{r,R}\right)$. It is
proportional to the mass ratio $\beta_{1}$.

The theory in this paper is based on a time dependent approach which can be
derived from an eikonal approximation. The projectile motion relative to the
target is described by a time-dependent classical trajectory $\mathbf{R}%
\left(  t\right)  =\mathbf{d}+vt\mathbf{\hat{z}}$ with constant velocity $v$
and impact parameter $\mathbf{d}$ ( $\mathbf{\hat{z}}$ is a unit vector
parallel to the z-axis). As discussed in Ref.\cite{mar02} the main effect of $V_{ct}(\mathbf{R})$ is to give an
absorption for small core-target impact parameters and thus it reduces the core survival probability.

Then  with the
approximations (\ref{vnt},\ref{vnc})  to the potentials the wave function
$\phi\left( \mathbf{r,d} ,t\right) $ describing the dynamics of the halo proton satisfies the
time-dependent equation
\begin{eqnarray}
i\hbar\frac{\partial\phi\left(  \mathbf{r,}\mathbf{d},t\right)  }{\partial
t}=(  H_{r}+V_{pt}(\mathbf{r+R}(  t) ) &+&V_{eff}(  \mathbf{r,R}(  t)  )  ) \nonumber \\ &\times& \phi(
\mathbf{r,}\mathbf{d},t)  \label{tde}
\end{eqnarray}
where $H_{r}=T_{r}+V_{pc}\left(  \mathbf{r}\right)  $ is the Hamiltonian for
the halo nucleus. In the present paper we neglect the nuclear part of final state interactions between the
proton and the halo core, but include the Coulomb proton-core interaction $V_{pc}(\mathbf{r})=\frac {Z_h Z_Ce^2}{|{\bf
r}|}$  and the final state interactions between the proton and the target.  This approximation should be satisfactory
unless there are resonances in the proton-core final state interaction which are strongly excited during the reaction.
 The proton-core potential does not act
dynamically and it cannot cause breakup. It gives the maximum contribution at the top of the proton-core barrier
 where $|{\bf r}|=R_i$. 
Therefore we take it
constant as $V_{pc}=\frac {Z_h Z_Ce^2}{R_i}$.

 When the nuclear proton-core final
state interactions are neglected 
we can define a potential \begin{eqnarray}\bar{V}_{2}(  \mathbf{r,}t)  &=&V^N_{pt}(\mathbf{r+R}(
t))+V^C_{pt}(\mathbf{r+R}(
t)) +V_{eff}(  \mathbf{r,R}(t) ) \nonumber \\ &=&V^N_{pt}(\mathbf{r+R})+ V_{\rm
Coul,}\end{eqnarray}
and $V^N_{pt}$ and $V^C_{pt}$ are the nuclear and Coulomb parts of the proton-target interaction.

Thus in the case of a proton breakup and for a heavy target, besides the proton target nuclear potential
it is necessary to include in the total Hamiltonian the proton-core, proton-target
and core-target Coulomb potentials.  The proton-target potential and the
core-target potential are included dynamically but the effect on the center
of mass has to be subtracted. We have then
\begin{equation}
V_{\rm Coul} = Z_T
e^2\left(\frac{Z_h}{|{\bf R}+\beta_2{\bf r}|}+\frac{Z_C}{|{\bf R}-\beta_1{\bf r}|}
-\frac{Z_C}{|{\bf R}|}-\frac{Z_h}{|{\bf R}|} \right)
\label{eq4}
\end{equation}
where charges and masses are : core ($A_C$,$Z_C$), halo ($A_h$,$Z_h$), target
($A_T$,$Z_T$). We used also two ratios : $\beta_1=A_h/A_P$ and
$\beta_2=A_c/A_P\approx 1$, with $A_P=A_C+A_h$.

Now we approximate $V_{\rm Coul}$ with something simpler. One approach is to make the dipole
expansion. This is quite good for the core-target recoil term because $\beta_1$ is small, but is less
good for the halo-target term because $\beta_2 \approx 1$. It would be good for large separations when
$R>>r$. Another possibility is to make a dipole approximation rather than a dipole expansion. 

It means
 making an approximation to $V_{\rm Coul}$ which is reasonable over the region of the projectile by
writing
\begin{equation}
V_{\rm Coul}({\bf r}, {\bf R})\approx V_{{\rm C} 0} + C^{(D)}\frac{{\bf r}\cdot {\bf R}}{|{\bf R}|^3}
\end{equation}
and choosing $V_{{\rm C} 0}$ and $C^{(D)}$  so that, when
$\bf r$ and $\bf R$ are alligned, the approximation fits the exact expression Eq.(\ref{eq4})
 at $r=\pm R_i$,
at the barrier tops on the left and right of the projectile (see Fig. 2).

Thus we put 
\begin{equation}
V_{{\rm C} 0}= \frac{1}{2}\left(V_{\rm Coul}(R_i, R) + V_{\rm Coul}(-R_i, R) \right)
\end{equation}
In other words $V_{{\rm C} 0}$ is the average of $V_{\rm Coul}$  at the points $r=\pm R_i$,   on
the left and right sides of the projectile (see Fig. 2). In this way the maximum contribution of the
proton-target Coulomb potential ${Z_h Z_T
e^2}/{|{\bf R}+\beta_2{\bf r}|}$ is taken into account to all orders. 

The constant 
$C^{(D)}$ is  chosen so that
\begin{equation}
C^{(D)}\frac{R_i}{R^2}= \frac{1}{2}\left(V_{\rm Coul}(R_i, R) - V_{\rm Coul}(-R_i, R) \right).
\end{equation}
In the limit when $R$ is very large the dipole expansion is good and we have $V_{{\rm C} 0}\approx 0$, 
$C^{(D)}\approx  \beta_1 Z_T Z_C e^2-\beta_2 Z_T Z_h e^2$. 
For smaller values of $R$ the approximate form of $V_{\rm Coul}$ fits the exact Coulomb
potential near the left and right barriers of the projectile. The constants $V_{{\rm C} 0}$ and  $C^{(D)}$ are 
dependent on
$R$ but in the calculations one would use the values at the point of closest approach in the path of relative motion.

\medskip
This approximation would have several consequences: 

\medskip
\noindent 1) There would be an effective binding energy for the initial state
\begin{equation}
{\tilde \varepsilon_i}= {\varepsilon}_i- \frac{Z_C e^2}{R_i}-V_{{\rm C} 0}\label{a11}
\end{equation}
This is  Eq.(\ref{1a}) of the text. 

\medskip
\noindent 2) Because the approximate form of $V_{\rm Coul}$ is a dipole  field the breakup can be calculated
by the methods used for neutron breakup.  The only differences would be the effective strengths
$C^{(D)}$  and the effective binding energy in the initial state ${\tilde \varepsilon_i}$.

\medskip
\noindent 3) The main conclusion of the present  discussion is  that the
effective binding modifies the halo character of proton breakup reactions. 

\medskip
 Then the initial condition that as
$t\rightarrow-\infty$ the wave function tends to the initial halo nucleus wave-function reads
\begin{equation}
\phi\left(  \mathbf{r,}\mathbf{d},t\right)  \rightarrow\phi_{lm}\left(
\mathbf{r},t\right)  =\phi_{lm}\left(  \mathbf{r}\right)  \exp\left(
-i\tilde \varepsilon_{i}t/\hbar\right)  \label{wf0}%
\end{equation} provided  the separation energy is given by Eq.(\ref{a11}).

 Now

\begin{equation}
\mathbf{\tilde V}_{eff}({\bf r},t) =C^{(D)}\frac{{\bf r}\cdot {\bf R}}{|{\bf R}|^3}\label{eq55}\end{equation}  
and the final proton wave function  $\phi_{f}\left(  t\right)  $ satisfies the equation%
\begin{equation}
i\hbar\frac{\partial\phi_{f}\left(  t\right)  }{\partial t}=(T_{r}+\tilde{V}%
_{2}\left(  \mathbf{r},t\right)  )\phi_{f}\left(  t\right)
\end{equation} with
 $$\tilde{V}_{2}\left(  \mathbf{r,}t\right)  =V^N_{pt}(\mathbf{r+R}\left(
t\right)  \mathbf{)+\tilde V}_{eff}\left(  \mathbf{r,R}\left(  t\right)  \right),  $$
  $V^N_{pt}$ would just be the nuclear part of the proton-target final state interaction and
 the boundary condition that $\phi_{f}\left(  t\right)  \sim\exp\left(
i\mathbf{k \cdot r}-\varepsilon_{f}t/\hbar\right)  $ when $t$ is large. 

The
final step is to make an eikonal approximation for $\phi_{f}\left(  t\right)
$
\begin{equation}
\phi_{f}\left(  t\right)  =\exp\left(  i\mathbf{k \cdot r}-i\varepsilon_{f}
t/\hbar\right)  \exp\left(  -\frac{1}{i\hbar}\int_{t}^{\infty}\tilde{V}%
_{2}\left(  \mathbf{r},t^{\prime}\right)  dt^{\prime}\right)  \label{wfe}%
\end{equation}
 such that we obtain   for the amplitude 

\begin{eqnarray}
\tilde g_{lm}\left(  \mathbf{k,}\mathbf{d}\right)  =&&\frac{1}{i\hbar}\int
d^{3}\mathbf{r}\int dte^{-i\mathbf{k \cdot r}+i\tilde\omega t}e^{\left(  \frac{1}{i\hbar
}\int_{t}^{\infty}\tilde{V}_{2}\left(  \mathbf{r},t^{\prime}\right)  dt^{\prime}\right)  }
 \nonumber \\ &&\times \tilde
{V}_{2}\left(  \mathbf{r},t\right)  \phi_{lm}\left(  \mathbf{r}\right)
\label{amp2}%
\end{eqnarray}
 with a new $\tilde \omega=\left(  \varepsilon_{f}-\tilde \varepsilon_{i}\right)  /\hbar.
$ 
Equation (\ref{amp2}) is formally the same as the neutron breakup amplitude of Ref.\cite{mar02} with the difference that
the effect of the halo charge has been included in the effective energy $\tilde \varepsilon_i$.

\end{document}